\documentclass{PoS}

\title{$\beta$-functions in higher dimensional field theories}

\ShortTitle{$\beta$-functions in higher dimensional field theories}

\author{\speaker{J.A. Gracey}\\
        Theoretical Physics Division, Department of Mathematical Sciences,
        University of Liverpool, P.O. Box 147, Liverpool, L69 3BX,
        United Kingdom \\
        E-mail: \email{gracey@liverpool.ac.uk}}

\abstract{We review recent activity in the construction of the renormalization 
group functions for $O(N)$ scalar and gauge theories in six and higher
dimensions. The theories lie in their respective universality classes at the
Wilson-Fisher fixed point. The critical exponents at this fixed point in the
various dimensions are all in agreement with the known exponents determined in
the large $N$ expansion.}

\FullConference{Loops and Legs in Quantum Field Theory\\
		24-29 April 2016\\
		Leipzig, Germany}

\newcommand{\Dslash}{D \! \! \! \! \! /}
\newcommand{\partialslash}{\partial \! \! \! \! \! /}

\newcommand{\MSbar}{\overline{\mbox{MS}}}

\newcommand{\Nf}{N_{\!f}}

\begin{document}

\section{Introduction}

Dimensional regularization is the canonical method of controlling the 
divergences when renormalizing a quantum field theory in perturbation theory.
The theory is extended from the critical dimension $D$ to $d$ where
$d$~$=$~$D$~$-$~$2\epsilon$ and $\epsilon$ is the regularizing parameter. The 
poles in $\epsilon$ are then absorbed in renormalization constants prior to the 
construction of the renormalization group functions. The process is complete 
when these functions are determined in $d$~$=$~$D$ dimensions. What is less 
apparent in the procedure is the connectivity different theories, with the same
underlying symmetry, have with each other in different dimensions. This 
connection derives from the critical point renormalization group equation and 
the Wilson-Fisher (WF) fixed point which is defined as the non-trivial zero of 
the $\beta$-function in $d$-dimensions. This connectivity has seen a recent 
revival in analysing conformal field theories beyond two dimensions as well as 
related interest in the $a$-theorem, ultraviolet-infrared duality across 
dimensions and conformal windows. Here we review the renormalization group 
functions of scalar and gauge theories in dimensions higher than four in their 
respective universality classes.

\section{Scalar theories}

We begin by reprising the picture of a tower of theories in $d$-dimensions all
lieing in the same universality class at the WF fixed point by considering 
scalar theories with an $O(N)$ symmetry. The base theory is in two dimensions 
which is the $O(N)$ nonlinear $\sigma$ model with the Lagrangian
\begin{equation}
L^{\mbox{\footnotesize{$\sigma$}}} ~=~
\frac{1}{2} \left( \partial_\mu \phi^i \right)^2 ~+~ \frac{1}{2} \sigma
\left( \phi^i \phi^i - \frac{1}{\lambda} \right) 
\label{lag2}
\end{equation}
where $\lambda$ is the coupling constant and $\sigma$ is a Lagrange multiplier
field. Throughout the set of theories in the same universality class the two
fields have dimensions $[\phi^i]$~$=$~$d/2-1$ and $[\sigma]$~$=$~$2$ in
$d$-dimensions. While (\ref{lag2}) is non-renormalizable perturbatively it is
renormalizable in the $1/N$ expansion which is a dimensionless coupling
parameter in $d$-dimensions. This point of view of the renormalizability is the
key to understanding the dimensional connectivity. The common feature of 
theories in the same universality class is the core interaction such as that in
(\ref{lag2}). For instance, relative to four dimensions (\ref{lag2}) is 
critically equivalent to $O(N)$ $\phi^4$ theory which can be formulated in two 
ways since 
\begin{equation}
L^{(4)} ~=~ \frac{1}{2} \left( \partial_\mu \phi^i \right)^2 ~-~ \frac{g}{8}
\left( \phi^i \phi^i \right)^2 ~=~
\frac{1}{2} \left( \partial_\mu \phi^i \right)^2 ~+~
\frac{1}{2} \sigma \phi^i \phi^i - \frac{\sigma^2}{2 g} ~.
\label{lag4}
\end{equation}
In the latter Lagrangian $\sigma$ is an auxiliary field and appears 
quadratically rather than linearly to ensure perturbative renormalizability in
the critical dimension. These additional $\sigma$ dependent parts of a 
Lagrangian over and above the core interaction are termed spectators since they
are only present in specific dimensions. Once (\ref{lag2}) and (\ref{lag4}) are
viewed in this way the method to construct the Lagrangians in the universal 
tower is evident and requires a core interaction and renormalizability. The six 
dimensional extension, $L^{(6)}$, was given in \cite{1,2,3,4,5} and extended to
eight in \cite{6} and their respective Lagrangians are  
\begin{eqnarray}
L^{(6)} &=& \frac{1}{2} \left( \partial_\mu \phi^i \right)^2 ~+~
\frac{1}{2} \left( \partial_\mu \sigma \right)^2 ~+~
\frac{g_1}{2} \sigma \phi^i \phi^i ~+~ \frac{g_2}{6} \sigma^3 \nonumber \\
L^{(8)} &=& \frac{1}{2} \left( \partial_\mu \phi^i \right)^2 ~+~
\frac{1}{2} \left( \Box \sigma \right)^2 ~+~
\frac{1}{2} g_1 \sigma \phi^i \phi^i ~+~
\frac{1}{6} g_2 \sigma^2 \Box \sigma ~+~ \frac{1}{24} g_3^2 \sigma^4 
\label{lag68}
\end{eqnarray}
where the number of independent couplings increases with dimension. For all of
these theories the critical exponents which are derived from the 
renormalization group functions have been computed in the large $N$ expansion, 
\cite{7,8,9,10}. For instance, the first three terms of the exponents
\begin{equation}
\eta ~=~ \gamma_\phi(g_c) ~~~,~~~ \omega ~=~ \beta^\prime(g_c)
\end{equation}
are known, \cite{7,8,9,10}, as functions of $d$ where $\gamma_\phi(g)$ is the 
wave function anomalous dimension and $g_c$ represents the vector of critical 
couplings. It is worth noting that the derivation of large $N$ exponents uses 
analytic regularization of the Feynman integrals and not dimensional 
regularization which is effectively the reason why one can study field theories
in $d$-dimensions. Moreover, the large $N$ method is applicable to non-abelian 
gauge theories but $\Nf$, the number of quark flavours, is used as the 
expansion parameter rather than the number of colours. 
 
Recent activity for $L^{(6)}$ has involved the extension of the three loop
results of \cite{11,12} to the $O(N)$ case, \cite{3,5}, and then to four loops,
\cite{13}. For the latter computation one can derive all the basic 
renormalization group functions from evaluating a $2$-point function. This is
because in six dimensions a $1/(k^2)^2$ propagator is infrared safe unlike in
four dimensions. Therefore to extract the coupling constant renormalization one
can nullify one external leg momentum on the vertex function relegating it 
effectively to a $2$-point function evaluation. Such a nullification can be
accommodated within the usual renormalization of the field $2$-point function, 
\cite{13}. This approach substantially reduces the number of Feynman
graphs to be evaluated. However, to achieve this we have used the Laporta
algorithm, \cite{14}, and specifically the {\sc Reduze} encoding of it,
\cite{15,16}, to construct all the integration by parts relations between the 
required integrals. The final step requires the substitution of the basic 
master integrals. As the $2$-point four loop master integrals are known in four
dimensions, \cite{17}, we can determine the corresponding ones in six 
dimensions by applying Tarasov's method, \cite{18,19}. We use {\sc Form},
\cite{20}, throughout to handle the underlying algebra. The main results are 
provided in \cite{13} but we note that the renormalization group functions for 
the $O(1)$ version of $L^{(6)}$ are, \cite{11,12,13},
\begin{eqnarray}
\beta(g) &=& \frac{3}{8} g^3 ~-~ \frac{125}{288} g^5 ~+~
5 [ 2592 \zeta_3 + 6617 ] \frac{g^7}{41472} \nonumber \\
&& +~ [ -~ 4225824 \zeta_3 + 349920 \zeta_4 + 1244160 \zeta_5 - 3404365 ]
\frac{g^9}{1492992} \nonumber \\
\gamma_\phi(g) &=& -~ \frac{1}{12} g^2 ~+~ \frac{13}{432} g^4 ~+~
[ 2592 \zeta_3 - 5195 ] \frac{g^6}{62208} \nonumber \\
&& +~ [ 10080 \zeta_3 + 18144 \zeta_4 - 69120 \zeta_5 + 53449 ]
\frac{g^8}{248832} 
\end{eqnarray}
where $\zeta_n$ is the Riemann zeta-function. Finally, all the renormalization
group functions evaluated for the $O(N)$ versions of (\ref{lag68}), 
\cite{2,3,6,13}, are in full agreement with the known large $N$ exponents which
supports the Wilson's vision of a tower of theories in $d$-dimensions within a 
universal underlying theory. 

\section{Gauge theories}

We can also extend the process to non-supersymmetric gauge theories. First, we
focus on six-dimensional QCD. The gauge fixed Lagrangian in a linear covariant
gauge is, \cite{6}, 
\begin{eqnarray}
L^{(6)}_{\mbox{\footnotesize{GI}}} &=&
-~ \frac{1}{4} \left( D_\mu G_{\nu\sigma}^a \right)
\left( D^\mu G^{a \, \nu\sigma} \right) ~+~
\frac{g_2}{6} f^{abc} G_{\mu\nu}^a \, G^{b \, \mu\sigma} \,
G^{c \,\nu}_{~~~\sigma} ~+~ i \bar{\psi}^{iI} \Dslash \psi^{iI} \nonumber \\
&& -~ \frac{1}{2\alpha} \left( \partial_\mu \partial^\nu A^a_\nu \right)
\left( \partial^\mu \partial^\sigma A^a_\sigma \right) ~-~
\bar{c}^a \Box \left( \partial^\mu D_\mu c \right)^a
\label{lagqcd6}
\end{eqnarray}
where $\alpha$ is the gauge parameter. In six dimensions there are two
independent gluonic operators, \cite{21}, as a consequence of the Bianchi
identities and we have chosen to use a $2$-leg and a $3$-leg operator. The
latter is important in effective field theories in four dimensions. Though in 
six dimensions a $4$-fermi operator has dimension $10$ and hence is absent in
(\ref{lagqcd6}). As a consequence the gluon and ghost propagators have double
poles in the squared momentum unlike in four dimensions. Structurally
(\ref{lagqcd6}) is similar to the eight dimensional $O(N)$ scalar theory,
\cite{6}. However, as an aside the propagators of (\ref{lagqcd6}) would lead to
a confining inter-quark potential in {\em four} dimensions. Indeed there is a 
Schwinger-Dyson solution of four dimensional QCD which constructs the effective
infrared Lagrangian, \cite{22}, and takes the precise form (\ref{lagqcd6}) with
$g_2$ massive. Continuing this theme (\ref{lagqcd6}) can be extended to include
lower dimension operators with 
\begin{eqnarray}
L_m^{(6)} &=& L^{(6)} + m_1 \bar{\psi}^{iI} \psi^{iI} ~-~
\frac{1}{4} m_2^2 G_{\mu\nu}^a G^{a \, \mu\nu} ~-~
\frac{1}{2\alpha} m_3^2 (\partial^\mu A^a_\mu)^2 \nonumber \\
&& -~ m_3^2 \bar{c}^a \left( \partial^\mu D_\mu c \right)^a ~-~
\frac{1}{2} m_4^4 A^a_\mu A^{a \, \mu} ~+~ m_4^4 \alpha \bar{c}^a c^a ~.
\label{lagqcd6m}
\end{eqnarray}
Here $m_i$ are masses to ensure each term is dimension six. Interestingly the 
Landau gauge propagators derived from (\ref{lagqcd6m}) are formally the same as
those which are used to model the infrared structure of the propagators on the 
lattice in four dimensions, \cite{23}. For instance,
\begin{eqnarray}
\left. \langle A^a_\mu(p) A^b_\nu(-p) \rangle \right|_{\alpha=0} &=& -~
\frac{\delta^{ab}}{[(p^2)^2 + m_2^2 p^2 + m_4^4 ]} \left[ \eta_{\mu\nu} ~-~
\frac{p_\mu p_\nu}{p^2} \right] \nonumber \\
\left. \langle c^a(p) \bar{c}^b(-p) \rangle \right|_{\alpha=0} &=& -~
\frac{\delta^{ab}}{p^2[p^2 + m_3^2 ]} ~.
\end{eqnarray}

Returning to the massless Lagrangian we have calculated the wave function
anomalous dimensions and the $\beta$-functions to two loops for non-zero
$\alpha$. For the latter we evaluated the three $3$-point vertices at the
fully symmetric point with non-zero momenta flowing through each external leg.
Unlike in a renormalization of four dimensional QCD nullifying an external leg
momentum is not infrared safe in six dimensions as such an operation would
produce propagators of the form $1/(k^2)^4$ where $k$ is a loop momentum. By
renormalizing these three vertices and obtaining the same $\MSbar$ results for
the renormalization of $g_1$ is a non-trivial check on our computation. Like
the scalar theory analysis we have again used the Laporta algorithm, \cite{14},
to obtain our renormalization group functions. For instance, the two $\MSbar$ 
$\beta$-functions are, \cite{6},  
\begin{eqnarray}
\beta_1(g_1,g_2) &=& \left[ - 249 C_A - 16 \Nf T_F \right] \frac{g_1^3}{120}
\nonumber \\
&& +~ \left[ - 50682 C_A^2 g_1^3 + 2439 C_A^2 g_1^2 g_2 + 3129 C_A^2 g_1 g_2^2
- 315 C_A^2 g_2^3
- 1328 C_A \Nf T_F g_1^3
\right. \nonumber \\
&& \left. ~~~~
- 624 C_A \Nf T_F g_1^2 g_2 + 96 C_A \Nf T_F g_1 g_2^2
- 3040 C_F \Nf T_F g_1^3 \right] \frac{g_1^2}{4320} \nonumber \\
\beta_2(g_1,g_2) &=& \left[ 81 C_A g_1^3 - 552 C_A g_1^2 g_2
+ 135 C_A g_1 g_2^2 - 15 C_A g_2^3 + 104 \Nf T_F g_1^3
- 48 \Nf T_F g_1^2 g_2 \right] \frac{1}{120} \nonumber \\ 
\label{qcdbeta6}
\end{eqnarray}
where for space reasons we only note the one loop term of $\beta_2(g_1,g_2)$.
The full expression is given in \cite{6}. We note that the gauge coupling
constant is asymptotically free for all (positive) values of $\Nf$. A final 
check on the calculation of the renormalization group functions is if the 
critical exponents derived from them at the WF fixed point agree with the large
$\Nf$ exponents computed using Vasiliev's method. In this instance the base 
theory in two dimensions which serves to seed this fixed point universal theory
is the non-abelian Thirring model (NATM) which can be written in two ways as
\begin{equation}
L^{\mbox{\footnotesize{NATM}}} ~=~ i \bar{\psi}^i \partialslash \psi^i ~+~
\frac{\tilde{g}}{2} \left( \bar{\psi}^i T^a \gamma^\mu \psi^i \right)^2 ~=~ 
i \bar{\psi}^i \Dslash \psi^i ~-~ \frac{1}{2} A^a_\mu A^{a \, \mu}
\label{lagnatm}
\end{equation}
where in strictly two dimensions 
$A^a_\mu$~$\propto$~$\bar{\psi}^i T^a \gamma_\mu \psi^i$ is an auxiliary field.
Various large $\Nf$ critical exponents are known, \cite{24,25,26}, and 
expanding these in an $\epsilon$-expansion relative to six dimensions the two 
loop renormalization group functions are in total agreement. That this happens 
for (\ref{lagqcd6}) indicates that it is correct which is not as remarkable as 
the observation of \cite{27} that (\ref{lagnatm}) does not contain gluonic 
vertices. The large $\Nf$ $d$-dimensional critical exponents derived from
(\ref{lagnatm}) contain information on the contribution from the triple, 
quartic and {\em quintic} vertices of the higher dimensional gauge theories. 

Evidence for the connection of the field theories across the dimensions is not 
just restricted to the basic renormalization group functions. We have also 
computed the two loop $\MSbar$ corrections to the anomalous dimensions, 
$\gamma_{(n)}(g_1,g_2)$, of the flavour non-singlet twist-$2$ Wilson operators 
$\bar{\psi} \gamma^{\mu_1} D^{\mu_2} \ldots D^{\mu_n} \psi$ using 
(\ref{lagqcd6}). For the lowest moments we find 
\begin{eqnarray}
\gamma_{(2)}(g_1,g_2) &=& 2 C_F g_1^2 \nonumber \\
&& + [ 26841 C_A g_1^2 - 1200 C_A g_1 g_2 - 600 C_A g_2^2 - 4200 C_F g_1^2
+ 1264 g_1^2 \Nf T_F ] \frac{C_F g_1^2}{1800} \nonumber \\
\gamma_{(3)}(g_1,g_2) &=& \frac{49}{15} C_F g_1^2 \nonumber \\
&& + [ 186321 C_A g_1^2 - 10950 C_A g_1 g_2 - 4200 C_A g_2^2
- 23564 C_F g_1^2 + 9104 g_1^2 \Nf T_F ] \frac{7 C_F g_1^2}{54000} ~. 
\nonumber \\
\end{eqnarray}
Again evaluating these at the WF fixed point to determine the critical exponent
they are in exact agreement with the leading order large $\Nf$ exponent of 
\cite{28}.

One application of (\ref{qcdbeta6}) is that the location of the conformal
window can be determined in an $\epsilon$ expansion with 
$d$~$=$~$6$~$-$~$2\epsilon$. In four dimensional $SU(3)$ QCD the window is at
$9$~$\leq$~$\Nf$~$\leq$~$16$, \cite{29}. By numerically solving
\begin{equation}
\beta_1(g_1,g_2) ~=~ \beta_2(g_1,g_2) ~=~ 0 ~~,~~
\frac{\partial \beta_1}{\partial g_1} \frac{\partial \beta_2}{\partial g_2}
~-~ \frac{\partial \beta_1}{\partial g_2}
\frac{\partial \beta_2}{\partial g_1} ~=~ 0
\end{equation}
we find one real solution labelled by $A$ which is, \cite{6}, 
\begin{equation}
N_{\!f\,(A))} ~=~ 2.797566 \frac{C_A}{T_F} ~+~
\left[ 2.198165 C_F - 3.432003 C_A \right] \frac{\epsilon}{T_F} ~+~ 
O(\epsilon^2)
\end{equation}
where we retain the four dimensional trace convention. For $SU(3)$ we have
\begin{equation}
\left. N_{\!f\,(A)} \right|_{SU(3)} ~=~ 16.785398 ~-~ 14.730246 \epsilon ~+~
O(\epsilon^2) ~.
\end{equation}
Unlike the situation with $O(N)$ scalar theories the boundary in strictly six
dimensions is the same as in four dimensions. 

Further insight into the structure of higher dimensional gauge theories can be
gained by focusing on the extension of Quantum Electrodynamics (QED) to six and
{\em eight} dimensions. This is because one can compute to higher order than in
the non-abelian case especially in eight dimensions where there are a
substantial number of $4$-gluon operators. The respective Lagrangians are, 
\cite{30,6},
\begin{eqnarray}
\left. L^{(6)} \right|_{U(1)} &=& 
-~ \frac{1}{4} \left( \partial_\mu F_{\nu\sigma} \right)
\left( \partial^\mu F^{\nu\sigma} \right) ~-~
\frac{1}{2\alpha} \left( \partial_\mu \partial^\nu A_\nu \right)
\left( \partial^\mu \partial^\sigma A_\sigma \right) ~+~
i \bar{\psi}^i \Dslash \psi^i \nonumber \\
\left. L^{(8)} \right|_{U(1)} &=& 
-~ \frac{1}{4} \left( \partial_\mu \partial_\nu
F_{\sigma\rho} \right)
\left( \partial^\mu \partial^\nu F^{\sigma\rho} \right) ~-~
\frac{1}{2\alpha} \left( \partial_\mu \partial^\nu A_\nu \right)
\left( \partial^\mu \partial^\sigma A_\sigma \right) \nonumber \\
&& +~ i \bar{\psi}^i \Dslash \psi^i ~+~
\frac{g_2^2}{32} F_{\mu\nu} F^{\mu\nu} F_{\sigma\rho} F^{\sigma\rho} ~+~
\frac{g_3^2}{8} F_{\mu\nu} F^{\mu\sigma} F_{\nu\rho} F^{\sigma\rho} 
\label{lagqedd}
\end{eqnarray}
where we have extended the gauge fixing sector to eight dimensions for a linear
covariant gauge. The renormalization group functions of each theory are known
to various loop orders, \cite{30,6}, including the electron mass anomalous 
dimensions. In all cases the critical exponents derived from these are in full 
agreement with the corresponding critical exponents to whatever order they are 
known in large $\Nf$, \cite{31,24,26}. This again substantiates the picture of 
a tower of theories connected via the WF fixed point. Several general features 
emerge in the QED analysis. First, we have verified the result of \cite{30} 
that six dimensional QED is asymptotically free. By contrast the eight 
dimensional theory is like its four dimensional partner. Under the assumption 
that there are no triple photon vertices in any QED formulation in higher 
dimensions then $D$-dimensional QED is asymptotically free in 
$D$~$=$~$2$~$+$~$4r$ dimensions for integer $r$~$\geq$~$1$. For both 
Lagrangians (\ref{lagqedd}) we have explicitly checked that the Ward-Takahashi 
identity holds. In addition a novel feature emerges for the electron anomalous 
dimension in a linear covariant gauge. In four dimensions the gauge parameter 
appears only in the one loop term, \cite{32,33}, and is absent thereafter in 
any explicit evaluation. The six and eight dimensional theories share the same 
property to three and two loops respectively. Therefore, if a rigorous proof of
this observation emerges it should be applicable to all dimensions.  

\section{Discussion}

We have reviewed recent work in six and higher dimensional scalar and gauge 
field theories and provided solid evidence for the picture of a tower of
connected theories in $d$-dimensions accessed by the WF fixed point. This 
extends from strictly two dimensions where conformal field theories give the 
foundation for the connection. Results in gauge theories show a similar vision 
but with the observation that structurally eight dimensional scalar theory is 
similar to six dimensional QCD. From the higher dimensional quantum field 
theory side the next stage is to extend loop computations to higher order to 
refine the fixed point structure as will as to gain more insight into the 
operators which drive any infrared fixed points in QCD in the context of the 
underlying universal theory. Form the computational point of view one question 
is whether there is a deeper connection of the Tarasov construction of relating
$d$- and $(d+2)$-dimensional Feynman integrals with the underlying field 
theories. In other words is there a way of proceeding more fundamentally via a 
path integral construction without having to make the connection at the 
renormalization group function level?

\section*{Acknowledgements} This work was carried out with the support of the
STFC Consolidated Grant ST/L000431/1.

\end{document}